\documentclass[twocolumn,float,floatfix,preprintnumbers,aps,prl,showpacs,footinbib,amsmath,amssymb,amsfonts,superscriptaddress,tightenlines]{revtex4}

\usepackage{graphicx}
\usepackage[usenames]{color}
\usepackage{hyperref}

\newcommand{\be}{\begin{equation}}
\newcommand{\ee}{\end{equation}}
\newcommand{\bea}{\begin{eqnarray}}
\newcommand{\eea}{\end{eqnarray}}

\newcommand{\dd}{\mathrm{d}}
\newcommand{\stf}[1]{{\langle #1 \rangle}}

\newcommand{\cT}{{\cal T}}
\newcommand{\nbb}{{\cal N}}

\newcommand{\mystar}[3]{{{\cal B}_{#1}[#2,#3]}}

\begin{document}
\title{Parameterization of temperature and spectral distortions in future CMB experiments}

\author{Cyril Pitrou}
%\email{pitrou@iap.fr}
\affiliation{Institut d'Astrophysique de Paris,
CNRS UMR 7095, UPMC Univ Paris 06, 98 bis Bd Arago, 75014 Paris,
France\\
Sorbonne Universit\'es, Institut Lagrange de Paris, 98 bis Bd Arago, 75014 Paris, France}
\pacs{95.30.Jx, 98.70.Vc, 98.80.-k, 98.80.Es}

\author{Albert Stebbins}
%\email{stebbins@fnal.gov}
\affiliation{Theoretical Astrophysics Group, Fermi National
  Accelerator Laboratory, P.O. Box 500, Batavia, IL 60510}

\date{\today}

%%%%%%%%%%%%%%%%%%%%%%%%%%%%%%%%%%%%%%%%%%%%%%%%%%%%%%%%%%%%%%%%%%%%%%%
\begin{abstract}
CMB spectral distortions are induced by Compton collisions with electrons. We review the various  schemes to characterize the anisotropic CMB with a non-Planckian spectrum.   We advocate using logarithmically averaged temperature moments as the preferred language to describe these spectral distortions, both for theoretical modeling and observations.  Numerical modeling is simpler, the moments are frame-independent, and in terms of scattering the mode truncation is exact.
\end{abstract}
\maketitle

%Structure of collision term.
%Argumenter pourquoi d'autres variables pas possibles et d'autres possibles

%%%%%%%%%%%%%%%%%%%%%%%%%%%%%%%%%%%%%%%%%%%%%%%%%%%%%%%%%%%%%%%%%%%%%%%

The Planck Surveyor's measurements of the cosmic microwave background (CMB) have opened a 
new era in the analysis and extraction of cosmological information from CMB  data.
The CMB spectrum, {\it i.e.} departures from a black-body radiation (BB), have not only been used 
to detect ``point'' sources from the SZ effect~\cite{PlanckSZCosmo}, but we now have angular maps of SZ and other foregrounds to work with  ~\cite{Planckysky}.  We are entering an era where the CMB itself, both the intensity and polarization patterns, and not even including foregrounds, should no longer be treated as two-dimensional (angle dependence) but rather three dimensional including the spectral (frequency) dependence.  Cosmological information from all three dimensions needs to be extracted optimally~\cite{Chluba:2013pya} and theoretical modeling of this third dimension must be done accurately.

The frequency dependent brightness temperature is a complete description of the CMB which evolves according the Boltzmann equation, and is a frame dependent quantity.   When departures from BB are measurable, then a local definition of the ``temperature anisotropy'', $\delta T$, becomes ambiguous.  Ideally one would like to decompose  into a truncated series of modes, reducing partial into simpler ordinary differential equations.  Here we review the different proposed decompositions and argue that the logarithmically averaged temperature moments (LAM) first used in Ref.~\cite{Goodman:1995dt} and further developed in Ref.~\cite{Stebbins2007}, is the best decomposition and should be used for future theoretical modeling.  These LAM can be added to the spectral templates used to fit the CMB plus foregrounds.  

For the two angular dimensions, the assumed statistical isotropy guarantees that the spectrum,
$C_\ell$,  the bi-spectrum, $B_{\ell,\ell',\ell''}$, $\cdots$ are sufficient two-point, three-point, 
$\cdots$ statistics.  These spectra are based on a spherical harmonic decomposition of the angular
dependence.  For the spectral (frequency) decomposition we argue that the LAM are the most 
appropriate compact representation of the CMB spectral  distortions because {\it i}) it  leads to a 
frame invariant description of spectral distortions, which is independent from our local velocity, and {\it ii}) it is the variable with which the non-linear numerical integration is the simplest and was thus 
recently made possible for several groups~\cite{Huang:2012ub,Pettinari:2013he}.

%(Compton scattering with recoil plus stimulated emission), then Bremsstrahlung and double Compton scattering \cite{Chluba:2011hw}.

\subsection{General formalism for the description of a spectrum}

{\it Temperature transform.} The distribution function of radiation is a function of the position in
space-time, the direction of propagation and the energy $E$ of
radiation. It is of the form $n(E,\dots)$ (here $\dots$ indicates all
the non-spectral dependence and is often omitted below). 
%When performing a perturbative
%expansion around a background Friedmann Lema\^itre (FL) space-time,
%the global redshifting of photons due to cosmological expansion can
%be removed by using the conformal energy $a E$ rather than the energy~\cite{Naruko:2013aaa}, where $a$
%is the FL scale factor, leading to simpler equations since the background
%distribution function expressed with the conformal energy is constant
%in space-time.  From now on, $E$ indicates the conformal energy.
In previous literature~\cite{1972JETP...35..643Z,1975ApJ...198..245C,1992ApJ...385..288S,Chluba:2004cn} the starting point for the description of the spectral dependence  is
to consider that the CMB spectrum is a superposition of BBR with
different temperatures, given by the distribution $p(T,\dots)$, such that
\be\label{DefTransformsT}
n(E,\dots)=\int_{0}^\infty \dd T p(T,\dots) \nbb\left(E\over T\right) 
%=\int_{0}^\infty \dd T p(T) \nbb\left(\frac{E}{T}\right) 
\ee
with $\nbb (x) \equiv1/(\exp(x)-1)$.  If $\int_0^\infty p(T) dT\ne1$ the distribution is said to be ``gray''.
Ref.~\cite{Stebbins2007} gives a full treatment of grayness and there it is shown that an initially
non-gray  distribution with only Compton-type interactions will remain non-gray.  Henceforth we 
consider only non-gray distributions (see Ref.~\cite{Ellis:2013cu} for
an example of a process inducing grayness). 
One can characterize the shape of the spectrum by the moments of the distribution $p(T)$. One thus defines
\be\label{Powerlawmoments}
\bar T_{(p)}\equiv \left(\int_0^\infty T^p p(T)\, \dd T\right)^{1\over p}
\,.
\ee
Different authors have concentrated on following only specific moments.  The most commonly used are the Rayleigh-Jeans temperature, 
$\bar T_{\rm RJ} \equiv \bar T_{(1)}$~\cite{Chluba:2004cn};
the number density temperature, 
$\bar T_{\rm n}    \equiv\bar T_{(3)}$~\cite{Pitrouysky,Naruko:2013aaa,PitrouyEBsky};
and the bolometric temperature 
$\bar T_{\rm b}    \equiv \bar T_{(4)}$~\cite{Pitrou:2010sn,Creminelli:2011sq,Huang:2012ub};
giving respectively the low frequency brightness, the number density of photons, and the energy 
density in photons. 
%Indeed, using Eq. (\ref{DefTransformsT}) and
%$\int_0^\infty x^p {\cal N}(x) = p! \zeta(p+1)$, we find
%$(p-1)!\zeta(p){T_{(p)}}^p \propto \int_0^\infty n(E) E^{p-1}\dd E$
%for $p\ge 2$. 
Indeed, using Eq. (\ref{DefTransformsT}) we find
${T_{(p)}}^p \propto \int_0^\infty n(E) E^{p-1}\dd E$ for $p\ge
2$. Note that if the distribution
function has a chemical potential, as in the case of a general Bose-Einstein
distribution, the low energy limit is then a constant ($[\exp(\mu/T) -1]^{-1}$), and it is thus impossible to describe such distribution as a
superposition of BBR like in Eq.~(\ref{DefTransformsT})  whose low
energy limit is $\propto \bar T_{\rm
  RJ}/E$. See however the appendix for the treatment of the case where
the effect of the chemical potential is negligible at low energy.

However in Ref.~\cite{Stebbins2007} an alternative description of
the spectral distortions was  proposed. At the basis of the formalism,
is the use of the variable $\cT\equiv\ln T$ (where a reference unit of
temperature is implicit) whose distribution is $q(\cT)\equiv T p(T)$. The logarithmically averaged temperature 
is then simply defined by
\be
\bar \cT \equiv \langle \cT \rangle
 \equiv \ln \bar T\,,\quad{\rm with}\quad \langle f  \rangle \equiv \int_{-\infty}^\infty \dd \cT
f(\cT)q(\cT).
\ee

\noindent {\it Spectral moments.} The spectral distortions are characterized by the moments (LAM) of $q(\cT)$:  the moments about 0, $\{\eta_p \}$; the central moments, $\{u_p\}$; and the  moments about a reference temperature, $\{d_p\}$, {\it i.e}
\be
\eta_p \equiv \langle \cT^p \rangle\,,\quad u_p \equiv
\langle (\cT-\bar \cT)^p \rangle\,,\quad d_p \equiv
\langle (\cT-\cT_0)^p \rangle 
\label{moments}
\ee
where $\cT_0 \equiv \ln T_0$ and $T_0$ is an arbitrary reference temperature, usually chosen close to the mean. By construction, $u_1=0$, and since the spectrum is non-gray
$\eta_0=d_0=u_0=1$.  Using $\cT = (\cT- \bar \cT) + \bar\cT$ and 
$\cT = (\cT- \cT_0) +\cT_0$, the moments of eq.~(\ref{moments}) are related by Leibniz-type 
relations
\begin{align}
&u_p = \mystar{p}{-\bar \cT}{\{\eta_k\}}= \mystar{p}{-d_1}{\{d_k\}}\label{ufromdandeta}\\
&\eta_p = \mystar{p}{\bar \cT}{\{u_k\}}\qquad d_p = \mystar{p}{d_1}{\{u_k\}} \label{dandetafromu}
\end{align}
where $\mystar{p}{A}{\{B_k\}}\equiv \sum_{m=0}^p{p\choose m} A^{p-m} B_m$. The meaning of the moments is clear as one can reconstruct the spectrum by
\be
n(E) = \sum_{m=0}^\infty \frac{d_m}{m!} D^m \nbb\left(\frac{E}{T_0}\right) =\sum_{m=0}^\infty \frac{u_m}{m!} D^m \nbb\left(\frac{E}{\bar T}\right)
\ee
where $D^m \nbb(x) \equiv (-1)^m \dd^m \nbb(x)/\dd \ln (x)^m$. Thus $\{d_m\}$ and $\{u_m\}$ are 
the coefficients of a generalized Fokker-Planck expansion around  $T_0$ and $\bar T$, respectively.  
The $u_p$ are frame independent, but this is not the case for the other types of 
moments~\cite{Stebbins2007}.   While this is an infinite expansion the lower order moments do not 
depend dynamically on the higher order moments (see below).  The  observed spectrum as a 
function of frequency and direction requires knowledge of the observer frame  because of the 
Doppler effect and associated aberration, so one must also know $\bar \cT$ and thus 
$d_1=\bar \cT -\cT_0$.  Truncation is useful because it is the lowest order moments which are 
the most evident: the 1st moment give the ``temperature perturbation''  
$\delta T/T=\delta \ln T=d_1$;  while the 2nd moment give the Compton $y$ distortion, 
$y={1\over2}u_2={1\over2}(d_2-d_1^2)$.  These two moments are the ones most relevant for 
current observations.
 
{\it Spectral moment of a polarized spectrum.} Linear polarization will be generated by Compton 
scattering and the previous formalism can be extended to describe the polarization\
spectrum~\cite{Stebbins2007}.   We use a $2\times2$ matrix-valued distribution function,
$n^{ab}(E)$, to describe both intensity and polarization 
(see e.g. Ref.~\cite{Tsagas:2007yx} for a review). The trace ($n$), the symmetric traceless part ($n^{\stf{ab}}$), and the antisymmetric part give, respectively, the intensity,  the linear polarization,  and the circular polarization.  The latter is not generated by Compton collisions so we set it to zero. 
We thus have $ n^{ab}(E) = \delta^{ab }n(E) +n^{\stf{ab}}(E)$.
%The usual Stokes parameters for the linear polarization can be extracted from this matrix valued
%distribution functions thanks to $Q = (n^{11}-n^{22})/2$ and $U = (n^{12}+n^{21})/2$. 
%When the angular dependence is decomposed onto spin-weighted
%spherical harmonics~\cite{1967JMP.....8.2155G}, the ${\cal E}$ and ${\cal B}$ modes of
%polarization can then be obtained, and it remains to describe the
%spectral dependence of these ${\cal E}$ and ${\cal B}$ modes.
% What about frame invariance?
A matrix-valued distribution of BBR $q^{ab}(\cT)$ is defined by
\be\label{DefTransformsPolarization}
n^{ab}(E)=\int_{-\infty}^\infty \dd \cT q^{ab}(\cT)\ \nbb\left(E\,e^{-\cT}\right) 
\ee
and its matrix-valued moments $\{d^{ab}_p\}, \{\eta^{ab}_p\}, \{u^{ab}_p\}$ can be generalized from 
the $\{d_p\}, \{\eta_p\}, \{u_p\}$, for which a trace and a symmetric traceless part can be defined. 
The relations (\ref{ufromdandeta},\ref{dandetafromu}) are then straightforwardly extended for linear 
polarization.
% to 
%\bea
%u^\stf{ab}_p &=& \mystar{p}{-\bar \cT}{\{\eta^\stf{ab}_k\}}= \mystar{p}{-\widetilde{\Theta}}{\{d^\stf{ab}_k\}}\label{ufromdandetaPolar}\\
%\eta^\stf{ab}_p &=& \mystar{p}{\bar \cT}{ \{u^\stf{ab}_k\}}\quad d^\stf{ab}_p = \mystar{p}{\widetilde{\Theta}}{\{u^\stf{ab}_k\}} \label{dandetafromuPolar}
%\eea

%%TODO write in compact notation.
From the structure of the Compton collision term, it can be shown~\cite{Stebbins2007} that 
$d_0^{\stf{ab}} =\eta_0^{\stf{ab}} = 0$ if initially so, but $u_1^{\stf{ab}}\neq 0$.  The set of variables 
for the polarized part is thus simply the set of $\{u_p^\stf{ab}\}_{p \ge 1}$ as they are frame 
independent. Compared to the intensity, the main difference is that there is no temperature to be 
defined for polarization, but there is the non-vanishing moment $u_1^\stf{ab}$ which is the dominant 
one.  A common misstatement or misunderstanding consists in treating this moment as a 
temperature perturbation, and to use the definition $\Theta^\stf{ab} \equiv u_1^\stf{ab}$, but strictly 
speaking, it is a pure spectral distortion, and as such frame independent. In 
Ref.~\cite{Naruko:2013aaa}, it is called the {\it ``temperature part''} of the polarization, as opposed 
to the primary spectral distortion $u_2^\stf{ab}=d_2^\stf{ab} - 2d_1 d_1^\stf{ab}$.
%TODO Here we have to argue why the $u_n$ is better that the $d_n$,
%again it is because of frame invariance, but this maybe needs to be
%more explicite since it is not discussed in Stebbins 2007.

{\it Extraction of moments from a measured spectrum.} In appendix A
of Ref.~\cite{Stebbins2007}, it is shown that, from a spectrum $n(E)$, the
moments $\{\eta_p\}$ can be obtained from
\begin{eqnarray}
\eta_p& =& \sum_{m=0}^p {p \choose m} \tilde \kappa_{p-m} I_p[n(E)] \\
I_p[n(E)] &=& \int_{-\infty}^\infty (\ln E)^p [D^2 n(E) - D n(E)] \dd \ln E \nonumber
\end{eqnarray}
and the central moments are deduced from (\ref{ufromdandeta}).  The
lowest order $\tilde \kappa_p$
are computed in \cite{Stebbins2007}.
%The first few coefficients are \mbox{$\tilde \kappa_0 = 2$}, 
%\mbox{$\tilde\kappa_1 = -0.5213228$}, \mbox{$\tilde \kappa_2 = -2.4238210$}, 
%\mbox{$\tilde\kappa_3= 6.1729862$}.
If the signal is sampled in several bands, and if the number of bands is large enough (as should be 
the case for future CMB experiments~\cite{Andre:2013afa}), then linear combinations of these 
signals with appropriate weights would be equivalent to the numerical integrations $I_p[n^{ab}(E)]$,
and would thus allow determination of the temperature and the first spectral moments. This method
is straightforwardly extended for the extraction of polarization moments $\{\eta_p^\stf{ab}\}$ 
and then the $\{u_p^\stf{ab}\}$. 
%However, it is not possible to extract the moments for a distribution
%function with a chemical potential. Indeed 

{\it Discussion on the choice of a set of variables.} It is clear that since the $\{u_p\}_{p\ge 2}$ are 
frame invariant they are good candidates to describe the spectral distortions. The use of $d_1$ for 
the temperature perturbation is then natural as it fits into this formalism. However, one might wonder
if this is the only set of variables with such appealing properties. Starting from the moments defined 
in (\ref{Powerlawmoments}), we can relate these to the $\{d_p\}$ and $\{u_p\}$ by
\be\label{Tpowern}
\langle T^p\rangle = (\bar T_{(p)})^p=T_0^p \sum_{m} \frac{p^m d_m}{m!}
= \bar T^p \sum_{m} \frac{p^m u_m}{m!} \,.
\ee
It appears clearly that, for a given $p$, the temperature $\bar T_{(p)}$ can be used to define a 
temperature perturbation and the moments
\be\label{MomentMn}
\Theta_{(p)} \equiv \bar T_{(p)}/T_0-1,  \,\,\, M_{(p),m} \equiv \langle \left(T-\bar T_{(p)}\right)^m \rangle \bar T_{(p)}^{-m}\,.
\ee
The $\{M_{(p),m}\}_{m\ge 2}$ would be as good as the $\{u_m\}_{m\ge 2}$ to describe
the spectral distortions, since they are obviously frame invariant as
they involve only an (infinite) sum of products of the $\{u_p\}$.
%GIve lowest order of such moments so as to get an idea.
%e.g. choosing bolometric p =4, and looking at lowest order moment n=2
In the next section, we argue that to decide which set of variables
should be used, one should examine the dynamical evolution, and choose the one which
has the simplest structure, and for which numerical integration is simplified.
%, and for which the free-streaming is easily
%integrated numerically. This means reducing a much as possible the
%non-linear couplings.

\subsection{Dynamical evolution of spectral moments}

{\it General form of the Boltzmann equation.} The general form of the Boltzmann equation is (again we omit the dependence
in $(E,\dots)$ for brevity)
\be\label{GenBoltzmann}
L^{ab}[n] \equiv \frac{{\cal D} n^{ab} }{{\cal D} \eta} + \frac{\dd \ln E}{\dd
  \eta}\frac{\partial n^{ab}}{\partial \ln E} = C^{ab}[n]
\ee
where the convective derivative ${\cal D}/{\cal D}\eta$ acts on all the dependence except the spectral dependence, and accounts for the
effect of free streaming. The collision term can also be described by
its moments $\{\eta_p^{C,\,ab}\}, \{u_p^{C,\,ab}\}, \{d_p^{C,\,ab}\}$ which are
related by relations similar to (\ref{ufromdandeta}) and
(\ref{dandetafromu}). In order to find the evolution of the
$\{u_p^{ab}\}$, it proves simpler to first derive from~(\ref{GenBoltzmann}) the evolution
of the $\{d_p^{ab}\}$, and we get
\be
\frac{{\cal D} d^{ab}_m}{{\cal D} \eta} = m d^{ab}_{m-1} \frac{\dd \ln E}{\dd
  \eta} + d^{C, \,ab}_m\,.
\ee 
So for the temperature perturbation, the trace of $m=1$:
\be\label{Eqd1}
\frac{{\cal D} d_1}{{\cal D} \eta} =  \frac{\dd \ln E}{\dd \eta} + d^C_1\,.
\ee
If the spectrum is initially non-gray, and radiation is only subject
to Compton scattering, it remains so and this property translates to
$d_0^{C,\,ab} =0$. The moments $\{d_p^{C,ab}\}$ can be read off the collision
term (see e.g. Ref.~\cite{Beneke:2010eg}), and as long as the thermal
effects are ignored (or treated separately from the Kompaneets equation~\cite{Kompaneets1957}), 
they are linear in the variables $\{d_p\}$ which describe the radiation spectrum. However, they
still couple non-linearly to the baryons bulk velocity~\cite{Tsagas:2007yx,Pitrou:2008hy}.

From the relations~(\ref{ufromdandeta}), one infers that
\bea\label{Boltzmannu}
\frac{{\cal D} u^{ab}_p}{{\cal D} \eta} &=& \sum_{m=1}^p {m \choose p} \left(-d_1\right)^{p-m}\left[d^{C,\,ab}_m - m d^{ab}_{m-1}
d^C_1\right]\nonumber\\
&=&u^{C,ab}_p - p  u_{p-1}^{ab} d_1^C\,.
%&=&u^{C,\,ab}_p
\eea
This system of equation is closed at any order $p$, since the
equation-of-motion for  $u^{ab}_p$ depends only on  $u^{ab}_{p'}$ for
$p'\le p$.  One can truncate this system of equations at any order with no approximation!

{\it Doppler, SZ effect and $y$-type distortion.} At 1st order one
needs only the temperature perturbation $d_1$ and
$u_1^\stf{ab}=d_1^\stf{ab}$.   At 2nd order, one adds the spectral distortions $u_2$ and 
$u_2^\stf{ab}$, and this distortion, known in this context as the non-linear kinetic SZ
effect~\cite{Pitrouysky,PitrouyEBsky}, is generated by the r.h.s. of~(\ref{Boltzmannu}) with $p=2$.  

The distortion generated by the thermal SZ effect~\cite{1969ApSS...4..301Z} is also captured by 
$u_2$ and the usual $y$ parameter associated with it is related by 
$y  \equiv{1\over2}u_2 ={1\over2}(d_2 -d_1^2)$.
A polarized $y$-type distortion can also be defined~\cite{1980MNRAS.190..413S,Naruko:2013aaa,PitrouyEBsky} and is related to the moments by
$y^{\stf{ab}}\equiv{1\over2}u^{\stf{ab}}_2 ={1\over2}(d_2^\stf{ab}- 2 d_1 d_1^\stf{ab})$.

{\it Structure of the numerics.} Eq.~(\ref{Boltzmannu}) shows that
\begin{enumerate}
\item spectral distortions are affected only by the collision term, as they remain unaffected by metric
perturbations (see also Refs.~\cite{Stebbins2007,Pitrouysky,Naruko:2013aaa});
\item metric perturbations, which enter through the redshifting term $ \dd \ln E/\dd \eta $ affect only 
the evolution of the temperature perturbation $d_1$, and more importantly do not couple
non-linearly with $d_1$ [Eq. (\ref{Eqd1})];
\item the
collision term for the evolution of $u_p^{ab}$ [the r.h.s of~(\ref{Boltzmannu})], contains only terms of 
the form $d_1^{p-k} u_k^{ab}$ with $k\leq p$ (see Ref.~\cite{Stebbins2007} for more details) 
multiplied by powers of the baryons bulk velocity. Therefore it restricts the non-linearities to products 
of at most $p$ factors of spectral moments, when considering the evolution of the moment of order 
$p$. N.B. for $p=1$ the collision term ($d_1^{C,ab}$) is linear in the moments.
%, and it holds up to any order of perturbations.
\end{enumerate}
Any other parameterization of the distortion based on the $M_{(p),n}$ defined in (\ref{MomentMn}) 
would conserve property (1). However, property (3) would be lost with the $M_{(p),n}$. The 
loss of this property is, in principle, not a serious problem for the numerical integration, since 
interactions are localized in time by the visibility function. However, this would lead to unnecessary 
complications when going to higher orders of perturbations and thus higher moments. Our first 
argument here is that the simplest is the best.

Our second argument is that property (2) is crucial for the numerical integration since redshifting 
effects are not localized in time. Indeed, by avoiding a non-linear coupling between the temperature 
perturbations and the metric perturbations, the numerical integration is made possible even at the 
non-linear level as it avoids coupling between the angular moments of the temperature 
perturbations with the metric perturbation~\cite{Huang:2012ub}. Finding a form of the 
Boltzmann equation that satisfies this property, was the key to a successful numerical integration at 
second order~\cite{Huang:2012ub,Pettinari:2013he}. With the present formalism, this property 
arises  naturally for the variable $d_1$. Metric perturbations would also affect the geodesic and lead 
to  time-delay and  lensing effects, but these can be treated 
separately~\cite{Hu:2001yq,ZhiqiFilippo}.   There would be 
of course other variables for which property (2) holds. For instance, defining 
$\tilde \Theta_{(p)} \equiv \ln (1+\Theta_{(p)})$, one obtains from~(\ref{Tpowern}) that the 
variables
\be
\widetilde{\Theta}_{(p)} = d_1+\frac{1}{p} \ln\Big(1 + \sum_{m\ge 2} \frac{p^m u_m}{m!}\Big)
\ee
obviously satisfy property (2) but not property (3). Up to second order in cosmological 
perturbations (neglecting $\{u_p\}_{p\ge 3}$) the definitions for the most common temperatures are 
related by $d_1\simeq \widetilde \Theta_{\rm n}-\frac{3}{2} u_2 \simeq  
\widetilde \Theta_{\rm b}-2 u_2\simeq\widetilde \Theta_{\rm RJ}- \frac{1}{2}u_2\,$.
As an illustration, if we consider the bolometric temperature perturbation $\Theta_{\rm b}$, then up 
to second order in cosmological perturbations, only $d_1$ and $u_2$ need to be kept, and one finds that $\Theta_{\rm b} \simeq d_1+ \frac{1}{2} d_1^2 + 2 u_2$, but 
$\widetilde{\Theta}_{\rm b} \simeq d_1 + 2 u_2$. This motivated the use of 
$\widetilde{\Theta}_{\rm b}$ instead of $\Theta_{\rm  b}$ in the final output of 
Ref.~\cite{Huang:2012ub}, since property (2) is satisfied for the former and not for the latter. 
Similarly, for the fractional perturbation to the energy density, one finds up to second order in cosmological perturbations 
$\Delta \simeq 4 [d_1 + 2d_1^2 + 2 u_2] $,
 and using $\tilde \Delta \equiv \ln(1+\Delta)$, we find 
 $\tilde \Delta \simeq 4 (d_1 + 2 u_2)= 4 \tilde \Theta_{\rm b}$. Again this motivated the use of
$\tilde \Delta$ instead of $\Delta$ in the intermediate numerics of Ref.~\cite{Huang:2012ub}, so as 
to keep property (2) satisfied. A final example can be made with the fractional
energy density perturbation of linear polarization. One finds 
$\Delta^\stf{ab} \simeq 4 [d_1^\stf{ab}(1+4 d_1) + 2 u_2^\stf{ab}]$, and the non-linear term
$d_1^\stf{ab}d_1$ will induce a non-linear coupling of the type
$d_1^\stf{ab} \dd \ln E / \dd \eta$ in the evolution equation of $\Delta^\stf{ab}$. However, using 
$\tilde \Delta^\stf{ab} \equiv \Delta^\stf{ab} (1- 4 d_1) $, this non-linear coupling
disappears~\cite{GuidoChristian} and property (2) is recovered. 
In all these three examples, property (2) can be restored with an
ad-hoc change of variable, but property (3) is not satisfied, due to the term in $u_2$ for the
first two examples, and due to the term $u_2^\stf{ab}$ for the last
one. It implies in particular that the evolution equation for the
lowest order moment in this description, {\it i.e.} their temperature perturbation, has a collision term which is not linear in the moments of radiation.

%TODO Check this above
%Find the best argumentation
%Talk about the brightness perturbation which was used in HUang. That
%would be a concrete example
% Say that if we add some y it does not matter. What matters is the tilde.
%TOdo comment on the easiness to find the collision term from these expression.
%This results is valid at any order. It is THE variable
%Discuss briefly for polarization as well
%So is it frame invariant???
%Optimal variables deduced \cite{Huang:2012ub,GuidoChristian}.

% Find a title for a subsection here.

{\it Conclusion.} The essential properties described above for the structure of dynamical equations 
are only met with the set of variables made of $d_1$, $\{u_p\}_{p \ge 2}$ and
$\{u_p^\stf{ab}\}_{p\ge 1}$. Furthermore, the moments which characterize the spectral distortions 
are frame independent and thus do not depend on our local
velocity. Only the angular dependence is affected by the choice of
frame due to aberration effects. We strongly recommend that 
these moments should be used to parameterize the CMB spectrum. Furthermore, the analysis of spectral 
distortions from thermal effects, and other processes~\cite{Chluba:2011hw} could be rephrased in 
this unified language.\\
{\it Acknowledgements.} C.P. thanks J. Chluba, C. Fidler, Z. Huang, S. Renaux-Petel,
G. Pettinari, J.-P. Uzan, and F. Vernizzi for fruitful discussions. This work was supported by French state funds managed by the ANR within the Investissements d'Avenir programme under reference ANR-11-IDEX-0004-02.   AS was supported by the DOE at Fermilab under Contract No. DE-AC02-07CH11359.

\subsection{Appendix}

\begin{figure}[!htb]
	\includegraphics*[width=\hsize]{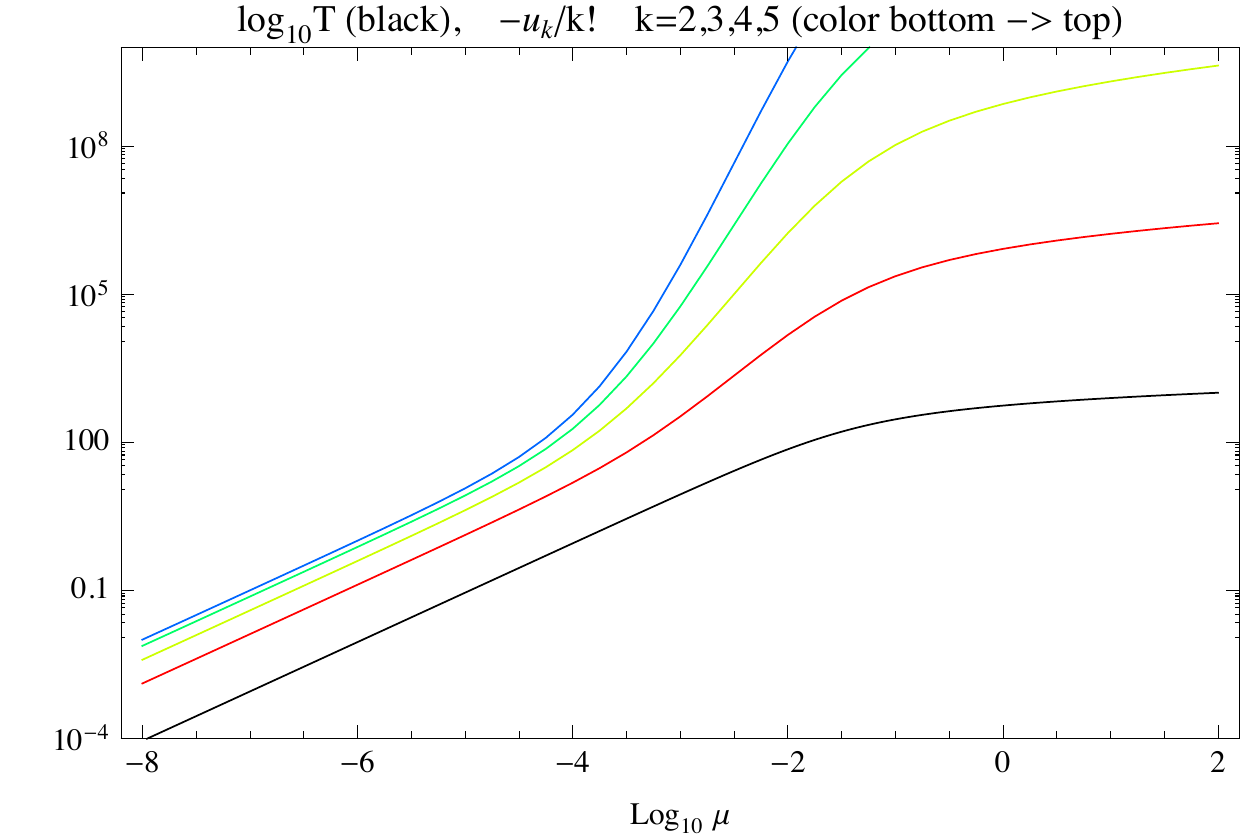}
\caption{The main moments for a pseudo Bose-Einstein distribution~(\ref{BEwithmucutoff})
  with cut-off $x_c=0.01$. In black : $\log_{10} \bar T$. In colors :
  $-u_k/k!$ for $k=2,3,4,5$ (from bottom to top).}
\label{fig1}
\end{figure}

{\it Distribution functions with a chemical potential.} At redshifts higher than $z \simeq 10^4$ and smaller than $z \simeq
3\,\times \, 10^{5}$, the thermalization of photons with an excess of energy with respect to a Planck spectrum results in a
Bose-Einstein distribution with a chemical
potential~\cite{Chluba:2013kua}. However, at low energies, processes such as double Compton emission and
Bremsstrahlung are enough to modify the spectrum and remove the effect
of the chemical potential so that the low energy limit of the
distribution is still $\propto 1/E$. It is indeed approximately described
by a Bose-Einstein distribution with an energy dependent chemical
potential. At high energies, this chemical potential converges to a
constant, but at low energies it is suppressed, and the transition
from one regime to the other is governed by a cut-off
$x_c=E_c/T$~\cite{Chluba:2013kua}. The distribution function of this ansatz is
approximately of the form
\be\label{BEwithmucutoff}
n(x) = \frac{1}{e^{x+\mu(x)} -1}\,,\quad \mu(x) = \mu_{\infty} e^{-x_c/x}\,.
\ee
with $x\equiv E/T$. A typical cut-off is $x_c=0.01$ and one can check that when
$\mu_\infty \to 0$, this distribution approaches a BBR as the
various moments decrease, as expected. In Fig.~\ref{fig1} we plot the first
moments as a function the chemical potential $\mu_\infty$ to
illustrate this convergence.

\bibliography{bibusky}

\begin{thebibliography}{29}
\expandafter\ifx\csname natexlab\endcsname\relax\def\natexlab#1{#1}\fi
\expandafter\ifx\csname bibnamefont\endcsname\relax
  \def\bibnamefont#1{#1}\fi
\expandafter\ifx\csname bibfnamefont\endcsname\relax
  \def\bibfnamefont#1{#1}\fi
\expandafter\ifx\csname citenamefont\endcsname\relax
  \def\citenamefont#1{#1}\fi
\expandafter\ifx\csname url\endcsname\relax
  \def\url#1{\texttt{#1}}\fi
\expandafter\ifx\csname urlprefix\endcsname\relax\def\urlprefix{URL }\fi
\providecommand{\bibinfo}[2]{#2}
\providecommand{\eprint}[2][]{\url{#2}}

\bibitem[{\citenamefont{Ade et~al.}(2013{\natexlab{a}})}]{PlanckSZCosmo}
\bibinfo{author}{\bibfnamefont{P.}~\bibnamefont{Ade}} \bibnamefont{et~al.}
  (\bibinfo{collaboration}{Planck Collaboration})
  (\bibinfo{year}{2013}{\natexlab{a}}), \eprint{1303.5080}.

\bibitem[{\citenamefont{Ade et~al.}(2013{\natexlab{b}})}]{Planckysky}
\bibinfo{author}{\bibfnamefont{P.}~\bibnamefont{Ade}} \bibnamefont{et~al.}
  (\bibinfo{collaboration}{Planck Collaboration})
  (\bibinfo{year}{2013}{\natexlab{b}}), \eprint{1303.5081}.

\bibitem[{\citenamefont{Chluba and Jeong}(2014)}]{Chluba:2013pya}
\bibinfo{author}{\bibfnamefont{J.}~\bibnamefont{Chluba}} \bibnamefont{and}
  \bibinfo{author}{\bibfnamefont{D.}~\bibnamefont{Jeong}},
  \bibinfo{journal}{MNRAS} \textbf{\bibinfo{volume}{438}},
  \bibinfo{pages}{2065} (\bibinfo{year}{2014}), \eprint{1306.5751}.

\bibitem[{\citenamefont{Goodman}(1995)}]{Goodman:1995dt}
\bibinfo{author}{\bibfnamefont{J.}~\bibnamefont{Goodman}},
  \bibinfo{journal}{Phys.Rev.D} \textbf{\bibinfo{volume}{52}},
  \bibinfo{pages}{1821} (\bibinfo{year}{1995}).

\bibitem[{\citenamefont{Stebbins}(2007)}]{Stebbins2007}
\bibinfo{author}{\bibfnamefont{A.}~\bibnamefont{Stebbins}}
  (\bibinfo{year}{2007}), \eprint{astro-ph/0703541}.

\bibitem[{\citenamefont{Huang and Vernizzi}(2013)}]{Huang:2012ub}
\bibinfo{author}{\bibfnamefont{Z.}~\bibnamefont{Huang}} \bibnamefont{and}
  \bibinfo{author}{\bibfnamefont{F.}~\bibnamefont{Vernizzi}},
  \bibinfo{journal}{Phys.Rev.Lett.} \textbf{\bibinfo{volume}{110}},
  \bibinfo{pages}{101303} (\bibinfo{year}{2013}).

\bibitem[{\citenamefont{Pettinari et~al.}(2013)\citenamefont{Pettinari, Fidler,
  Crittenden, Koyama, and Wands}}]{Pettinari:2013he}
\bibinfo{author}{\bibfnamefont{G.~W.} \bibnamefont{Pettinari}},
  \bibinfo{author}{\bibfnamefont{C.}~\bibnamefont{Fidler}},
  \bibinfo{author}{\bibfnamefont{R.}~\bibnamefont{Crittenden}},
  \bibinfo{author}{\bibfnamefont{K.}~\bibnamefont{Koyama}}, \bibnamefont{and}
  \bibinfo{author}{\bibfnamefont{D.}~\bibnamefont{Wands}},
  \bibinfo{journal}{JCAP} \textbf{\bibinfo{volume}{1304}}, \bibinfo{pages}{003}
  (\bibinfo{year}{2013}).

\bibitem[{\citenamefont{{Zel'dovich} et~al.}(1972)\citenamefont{{Zel'dovich},
  {Illarionov}, and {Syunyaev}}}]{1972JETP...35..643Z}
\bibinfo{author}{\bibfnamefont{Y.~B.} \bibnamefont{{Zel'dovich}}},
  \bibinfo{author}{\bibfnamefont{A.~F.} \bibnamefont{{Illarionov}}},
  \bibnamefont{and} \bibinfo{author}{\bibfnamefont{R.~A.}
  \bibnamefont{{Syunyaev}}}, \bibinfo{journal}{Soviet JETP}
  \textbf{\bibinfo{volume}{35}}, \bibinfo{pages}{643} (\bibinfo{year}{1972}).

\bibitem[{\citenamefont{{Chan} and {Jones}}(1975)}]{1975ApJ...198..245C}
\bibinfo{author}{\bibfnamefont{K.~L.} \bibnamefont{{Chan}}} \bibnamefont{and}
  \bibinfo{author}{\bibfnamefont{B.~J.~T.} \bibnamefont{{Jones}}},
  \bibinfo{journal}{Ap.J.} \textbf{\bibinfo{volume}{198}}, \bibinfo{pages}{245}
  (\bibinfo{year}{1975}).

\bibitem[{\citenamefont{{Salas}}(1992)}]{1992ApJ...385..288S}
\bibinfo{author}{\bibfnamefont{L.}~\bibnamefont{{Salas}}},
  \bibinfo{journal}{Ap.J.} \textbf{\bibinfo{volume}{385}}, \bibinfo{pages}{288}
  (\bibinfo{year}{1992}).

\bibitem[{\citenamefont{Chluba and Sunyaev}(2003)}]{Chluba:2004cn}
\bibinfo{author}{\bibfnamefont{J.}~\bibnamefont{Chluba}} \bibnamefont{and}
  \bibinfo{author}{\bibfnamefont{R.}~\bibnamefont{Sunyaev}},
  \bibinfo{journal}{A.\&A.} \textbf{\bibinfo{volume}{424}},
  \bibinfo{pages}{389} (\bibinfo{year}{2003}).

\bibitem[{\citenamefont{Ellis et~al.}(2013)\citenamefont{Ellis, Poltis, Uzan,
  and Weltman}}]{Ellis:2013cu}
\bibinfo{author}{\bibfnamefont{G.}~\bibnamefont{Ellis}},
  \bibinfo{author}{\bibfnamefont{R.}~\bibnamefont{Poltis}},
  \bibinfo{author}{\bibfnamefont{J.-P.} \bibnamefont{Uzan}}, \bibnamefont{and}
  \bibinfo{author}{\bibfnamefont{A.}~\bibnamefont{Weltman}},
  \bibinfo{journal}{Phys.Rev.D} \textbf{\bibinfo{volume}{87}},
  \bibinfo{pages}{103530} (\bibinfo{year}{2013}).

\bibitem[{\citenamefont{Pitrou et~al.}(2010{\natexlab{a}})\citenamefont{Pitrou,
  Bernardeau, and Uzan}}]{Pitrouysky}
\bibinfo{author}{\bibfnamefont{C.}~\bibnamefont{Pitrou}},
  \bibinfo{author}{\bibfnamefont{F.}~\bibnamefont{Bernardeau}},
  \bibnamefont{and} \bibinfo{author}{\bibfnamefont{J.-P.} \bibnamefont{Uzan}},
  \bibinfo{journal}{JCAP} \textbf{\bibinfo{volume}{1007}}, \bibinfo{pages}{019}
  (\bibinfo{year}{2010}{\natexlab{a}}).

\bibitem[{\citenamefont{Naruko et~al.}(2013)\citenamefont{Naruko, Pitrou,
  Koyama, and Sasaki}}]{Naruko:2013aaa}
\bibinfo{author}{\bibfnamefont{A.}~\bibnamefont{Naruko}},
  \bibinfo{author}{\bibfnamefont{C.}~\bibnamefont{Pitrou}},
  \bibinfo{author}{\bibfnamefont{K.}~\bibnamefont{Koyama}}, \bibnamefont{and}
  \bibinfo{author}{\bibfnamefont{M.}~\bibnamefont{Sasaki}},
  \bibinfo{journal}{Class.Quant.Grav.} \textbf{\bibinfo{volume}{30}},
  \bibinfo{pages}{165008} (\bibinfo{year}{2013}).

\bibitem[{\citenamefont{Renaux-Petel et~al.}(2014)\citenamefont{Renaux-Petel,
  Fidler, Pitrou, and Pettinari}}]{PitrouyEBsky}
\bibinfo{author}{\bibfnamefont{S.}~\bibnamefont{Renaux-Petel}},
  \bibinfo{author}{\bibfnamefont{C.}~\bibnamefont{Fidler}},
  \bibinfo{author}{\bibfnamefont{C.}~\bibnamefont{Pitrou}}, \bibnamefont{and}
  \bibinfo{author}{\bibfnamefont{G.~W.} \bibnamefont{Pettinari}},
  \bibinfo{journal}{JCAP} \textbf{\bibinfo{volume}{1403}}, \bibinfo{pages}{033}
  (\bibinfo{year}{2014}), \eprint{1312.4448}.

\bibitem[{\citenamefont{Pitrou et~al.}(2010{\natexlab{b}})\citenamefont{Pitrou,
  Uzan, and Bernardeau}}]{Pitrou:2010sn}
\bibinfo{author}{\bibfnamefont{C.}~\bibnamefont{Pitrou}},
  \bibinfo{author}{\bibfnamefont{J.-P.} \bibnamefont{Uzan}}, \bibnamefont{and}
  \bibinfo{author}{\bibfnamefont{F.}~\bibnamefont{Bernardeau}},
  \bibinfo{journal}{JCAP} \textbf{\bibinfo{volume}{1007}}, \bibinfo{pages}{003}
  (\bibinfo{year}{2010}{\natexlab{b}}).

\bibitem[{\citenamefont{Creminelli et~al.}(2011)\citenamefont{Creminelli,
  Pitrou, and Vernizzi}}]{Creminelli:2011sq}
\bibinfo{author}{\bibfnamefont{P.}~\bibnamefont{Creminelli}},
  \bibinfo{author}{\bibfnamefont{C.}~\bibnamefont{Pitrou}}, \bibnamefont{and}
  \bibinfo{author}{\bibfnamefont{F.}~\bibnamefont{Vernizzi}},
  \bibinfo{journal}{JCAP} \textbf{\bibinfo{volume}{1111}}, \bibinfo{pages}{025}
  (\bibinfo{year}{2011}).

\bibitem[{\citenamefont{Tsagas et~al.}(2008)\citenamefont{Tsagas, Challinor,
  and Maartens}}]{Tsagas:2007yx}
\bibinfo{author}{\bibfnamefont{C.~G.} \bibnamefont{Tsagas}},
  \bibinfo{author}{\bibfnamefont{A.}~\bibnamefont{Challinor}},
  \bibnamefont{and} \bibinfo{author}{\bibfnamefont{R.}~\bibnamefont{Maartens}},
  \bibinfo{journal}{Phys.Rept.} \textbf{\bibinfo{volume}{465}},
  \bibinfo{pages}{61} (\bibinfo{year}{2008}).

\bibitem[{\citenamefont{Andre et~al.}(2013)}]{Andre:2013afa}
\bibinfo{author}{\bibfnamefont{P.}~\bibnamefont{Andre}} \bibnamefont{et~al.}
  (\bibinfo{collaboration}{PRISM Collaboration}) (\bibinfo{year}{2013}).

\bibitem[{\citenamefont{Beneke and Fidler}(2010)}]{Beneke:2010eg}
\bibinfo{author}{\bibfnamefont{M.}~\bibnamefont{Beneke}} \bibnamefont{and}
  \bibinfo{author}{\bibfnamefont{C.}~\bibnamefont{Fidler}},
  \bibinfo{journal}{Phys.Rev.D} \textbf{\bibinfo{volume}{82}},
  \bibinfo{pages}{063509} (\bibinfo{year}{2010}).

\bibitem[{\citenamefont{Kompaneets}(1957)}]{Kompaneets1957}
\bibinfo{author}{\bibfnamefont{A.}~\bibnamefont{Kompaneets}},
  \bibinfo{journal}{JETP} \textbf{\bibinfo{volume}{4}}, \bibinfo{pages}{730}
  (\bibinfo{year}{1957}).

\bibitem[{\citenamefont{Pitrou}(2009)}]{Pitrou:2008hy}
\bibinfo{author}{\bibfnamefont{C.}~\bibnamefont{Pitrou}},
  \bibinfo{journal}{Class.Quant.Grav} \textbf{\bibinfo{volume}{26}},
  \bibinfo{pages}{065006} (\bibinfo{year}{2009}).

\bibitem[{\citenamefont{{Zeldovich} and {Sunyaev}}(1969)}]{1969ApSS...4..301Z}
\bibinfo{author}{\bibfnamefont{Y.~B.} \bibnamefont{{Zeldovich}}}
  \bibnamefont{and} \bibinfo{author}{\bibfnamefont{R.~A.}
  \bibnamefont{{Sunyaev}}}, \bibinfo{journal}{Astrophy. Sp. Sci.}
  \textbf{\bibinfo{volume}{4}}, \bibinfo{pages}{301} (\bibinfo{year}{1969}).

\bibitem[{\citenamefont{Sunyaev and Zeldovich}(1980)}]{1980MNRAS.190..413S}
\bibinfo{author}{\bibfnamefont{R.~A.} \bibnamefont{Sunyaev}} \bibnamefont{and}
  \bibinfo{author}{\bibfnamefont{I.~B.} \bibnamefont{Zeldovich}},
  \bibinfo{journal}{MNRAS} \textbf{\bibinfo{volume}{190}}, \bibinfo{pages}{413}
  (\bibinfo{year}{1980}).

\bibitem[{\citenamefont{Hu and Cooray}(2001)}]{Hu:2001yq}
\bibinfo{author}{\bibfnamefont{W.}~\bibnamefont{Hu}} \bibnamefont{and}
  \bibinfo{author}{\bibfnamefont{A.}~\bibnamefont{Cooray}},
  \bibinfo{journal}{Phys.Rev.D} \textbf{\bibinfo{volume}{63}},
  \bibinfo{pages}{023504} (\bibinfo{year}{2001}).

\bibitem[{\citenamefont{Huang and Vernizzi}(2014)}]{ZhiqiFilippo}
\bibinfo{author}{\bibfnamefont{Z.}~\bibnamefont{Huang}} \bibnamefont{and}
  \bibinfo{author}{\bibfnamefont{F.}~\bibnamefont{Vernizzi}},
  \bibinfo{journal}{Phys.Rev.} \textbf{\bibinfo{volume}{D89}},
  \bibinfo{pages}{021302} (\bibinfo{year}{2014}), \eprint{1311.6105}.

\bibitem[{\citenamefont{Fidler et~al.}(2014)\citenamefont{Fidler, Pettinari,
  Beneke, Crittenden, Koyama et~al.}}]{GuidoChristian}
\bibinfo{author}{\bibfnamefont{C.}~\bibnamefont{Fidler}},
  \bibinfo{author}{\bibfnamefont{G.~W.} \bibnamefont{Pettinari}},
  \bibinfo{author}{\bibfnamefont{M.}~\bibnamefont{Beneke}},
  \bibinfo{author}{\bibfnamefont{R.}~\bibnamefont{Crittenden}},
  \bibinfo{author}{\bibfnamefont{K.}~\bibnamefont{Koyama}},
  \bibnamefont{et~al.}, \bibinfo{journal}{JCAP} \textbf{\bibinfo{volume}{07}},
  \bibinfo{pages}{011} (\bibinfo{year}{2014}), \eprint{1401.3296}.

\bibitem[{\citenamefont{Chluba and Sunyaev}(2012)}]{Chluba:2011hw}
\bibinfo{author}{\bibfnamefont{J.}~\bibnamefont{Chluba}} \bibnamefont{and}
  \bibinfo{author}{\bibfnamefont{R.}~\bibnamefont{Sunyaev}},
  \bibinfo{journal}{MNRAS} \textbf{\bibinfo{volume}{419}},
  \bibinfo{pages}{1294} (\bibinfo{year}{2012}), \eprint{1109.6552}.

\bibitem[{\citenamefont{Chluba}(2014)}]{Chluba:2013kua}
\bibinfo{author}{\bibfnamefont{J.}~\bibnamefont{Chluba}},
  \bibinfo{journal}{MNRAS} \textbf{\bibinfo{volume}{440}},
  \bibinfo{pages}{2544} (\bibinfo{year}{2014}), \eprint{1312.6030}.

\end{thebibliography}

\end{document}